\documentclass[utf8]{FrontiersinHarvard}
\usepackage{url,hyperref,lineno,microtype,subcaption}
\usepackage[onehalfspacing]{setspace}

\def\keyFont{\fontsize{8}{11}\helveticabold }
\def\firstAuthorLast{Howell {et~al.}} 
\def\Authors{
Steve B. Howell\,$^{1}$,
Clara~E.~Martínez-Vázquez\,$^{2}$, 
Elise Furlan\,$^{3}$, 
Nicholas~J.~Scott\,$^{4}$, 
Rachel A. Matson\,$^{5}$, 
Colin Littlefield\,$^{6,1}$, 
Catherine~A.~Clark\,$^{3}$, 
Kathryn V. Lester\,$^{7}$, 
Zachary D. Hartman\,$^{1}$, 
David R. Ciardi\,$^{3}$,
Sarah J. Deveny\,$^{6,1}$
}
\def\Address{$^{1}$NASA Ames Research Center, Moffett Field, CA 94035, USA \\ 
$^{2}$ International Gemini Observatory/NSF NOIRLab, 670 N. A'ohoku Place, Hilo, HI 96720, USA\\
$^{3}$ NASA Exoplanet Science Institute, Caltech/IPAC, Mail Code 100-22, 1200 E. California Blvd., Pasadena, CA 91125, USA\\
$^{4}$ The CHARA Array, Mt. Wilson, CA 91023, USA\\
$^{5}$ U.S. Naval Observatory, 3450 Massachusetts Avenue NW, Washington, D.C. 20392, USA\\
$^{6}$ Bay Area Environmental Research Institute, Moffett Field, CA 94035, USA \\ 
$^{7}$ Mount Holyoke College, South Hadley MA 01075, USA
}

\def\corrAuthor{Steve B. Howell}

\def\corrEmail{steve.b.howell@nasa.gov}

\def\aj{{AJ}}                   
             
\def\apj{{ApJ}}                 
\def\apjl{{ApJ}}                
               
\def\ao{{Appl.~Opt.}}           
             
\def\aap{{A\&A}}

\def\mnras{{MNRAS}}

\def\pasp{{PASP}}

\begin{document}
\onecolumn
\firstpage{1}

\title[Nearly a Decade of Speckle Interferometry at Gemini]{Nearly a Decade of Groundbreaking Speckle Interferometry at the International Gemini Observatory}

\author[\firstAuthorLast ]{\Authors} %This field will be automatically populated
\address{} %This field will be automatically populated
\correspondance{} %This field will be automatically populated

\extraAuth{}% If there are more than 1 corresponding author, comment this line and uncomment the next one.
%\extraAuth{corresponding Author2 \\ Laboratory X2, Institute X2, Department X2, Organization X2, Street X2, City X2 , State XX2 (only USA, Canada and Australia), Zip Code2, X2 Country X2, email2@uni2.edu}

\maketitle

\begin{abstract}

\section{}
Since its inception, speckle interferometry has revolutionized high-resolution astronomical imaging, overcoming atmospheric challenges to achieve the diffraction limits of telescopes. Almost a decade ago, in 2018, a pair of speckle cameras -- `Alopeke and Zorro -- were installed at two of the largest apertures in the world, the twin 8.1-meter Gemini North and South telescopes in Hawai'i and Chile. Equipped with dual blue and red channels, 'Alopeke and Zorro deliver high-resolution imaging across the optical bandpass from 350 to 1000 nm, which has led to crucial discoveries in both stellar multiplicity and exoplanetary science. Furthermore, the broad and nonrestrictive access to these instruments, given by each Gemini Observatory partner and via the US NOIRLab open skies policy, has allowed the community to expand the applications of the instruments, supporting a wide range of scientific investigations from Solar System bodies, to morphological studies of stellar remnants and quasars, to evolved stars, to transient phenomena. This paper reviews the instrument technology and observational capabilities, and highlights key scientific contributions and discoveries of `Alopeke and Zorro, emphasizing the enduring importance of speckle interferometry in advancing modern observational astronomy and expanding the frontiers of astronomical research.

\tiny
 \keyFont{ \section{Keywords: Binary Stars, Planet Hosting Stars, Speckle Interferometry, Astronomical Techniques, High Angular Resolution}}
\end{abstract}

\section{Introduction} \label{sec:Intro}

Speckle interferometry, as a technique for high-resolution optical imaging, began in 1970 with the work of \citet{labeyrie:1970}, using ideas for fast imaging to remove atmospheric effects.
\citet{labeyrie:1970} showed that taking short exposures removed the effects of seeing-induced fluctuations that were causing distortions in the wavefront from a distant star. Removing these distortions allowed the diffraction limit of the telescope to be reached.

Speckle cameras initially used photographic plates, but continued with the various detectors of the day, such as photomultiplier tubes, video tubes, and reticons \citep[e.g.,][]{bonneau:1980,mcalister:1987,weigelt:1985,horch:1992,balega:1993}.  Initial studies mainly focused on bright binary stars using 1- to 4-meter telescopes and repeated imaging to produce precise stellar orbits \citep[e.g.][]{mcalister:1989}. Significant advances in astronomical detectors, in particular the more quantum efficient charge-coupled devices (CCDs), provided the next leap forward in this field, allowing for photon intensification, digital outputs, and higher signal-to-noise (S/N) observations to be obtained \citep[e.g.,][]{mcalister:1989,mason:1997,hartkopf:2000,horch:2000}. In recent years, the introduction of the electron-multiplying CCD (EMCCD) as a detector \citep{tokovinin:2008,horch:2011b} has been a game-changer. With ultra-fast readout, essentially zero read noise, near-perfect quantum efficiency, optical flatness, and ease of use, EMCCDs have revolutionized the field of speckle imaging. Fourier-based data reduction and image reconstructions using autocorrelation and power spectra techniques \citep{horch:2012,horch:2015} are enhanced using the bispectrum technique developed by \citet{Weigelt1977:OptCo..21...55W} and \citet{lohmann:1983} allowing phase information to be determined that resolves the ambiguity present in autocorrelation analysis \citep[e.g.,][]{howell:2011,hope:2022}.

This renaissance in speckle imaging has led to the development of new dedicated speckle instruments \citep[e.g.,][]{Mak2009AstBu..64..296M,toko2010PASP..122.1483T,Review-2021FrASS...8...10H,clark:2020,pedichini:2016} placed on some of the largest telescopes in the world. A summary of former and current speckle imagers in astronomy is presented in \citet{scott:2021}. Speckle imaging is no longer limited to bright star astrometry; it has expanded into many areas of point source and non-point source imagery \citep[e.g.,][]{Ricardo2020RNAAS...4..143S,scott:2021,SHARA2022MNRAS.509.2897S}. Fainter astronomical targets can now be observed \citep{howell:2021c}, the overall data quality and S/N ratio of the observations are greater, and the final reconstructed images have more fidelity \citep[e.g.,][]{2022FrASS...9.1163H}.

In contrast to infrared (IR) adaptive optics (AO) systems, optical speckle imaging on 8-meter-class telescopes routinely achieves an inner working angle (IWA) at the diffraction limit of the telescope \citep[20-30 mas across the optical][]{lester:2021}, uses far less expensive instrumentation, and does not require a (laser) guide star, enabling higher observational efficiency.
In speckle interferometry, the detection contrast is proportional to 1/(seeing)$^2$ but the angular resolution is unaffected ($\lambda$/D). This is in contrast to normal imaging and IR/AO in which the final image resolution is affected by the native seeing, being proportional to $\lambda$/r$_0$ \citep[r$_0$ $\sim$native seeing,][]{1966JOSA...56.1380F}. Speckle imaging performed in the optical bandpass (350 -1000 nm) provides the highest angular resolution available today on any single telescope, delivering $\sim$4 times better angular resolution than IR AO observations in the $K$-band.

This paper presents a summary of the first eight years of astronomical imaging observations using the highest resolution, deepest contrast speckle instruments available, `Alopeke and Zorro, which are mounted on the twin 8.1-meter Gemini North and South telescopes in Hawai'i and Chile \citep{scott:2021,2022FrASS...9.1163H}. We review the major advances and scientific areas covered with these instruments since their introduction in 2018.

\section{Zorro and `Alopeke: visiting speckle instruments at Gemini} \label{sec:Inst}

\begin{figure*}
\centering
\includegraphics[width=0.40\textwidth]{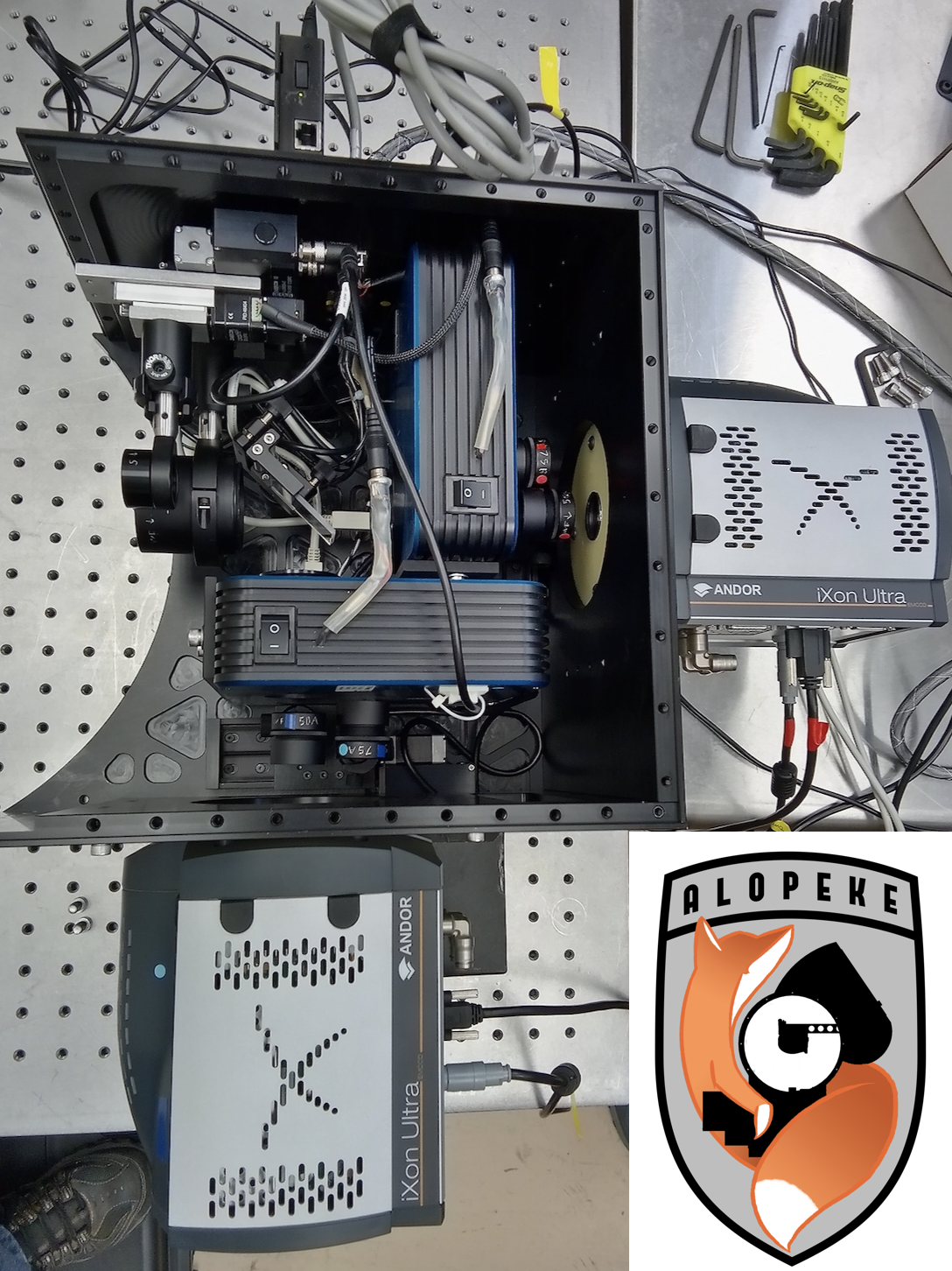}
\includegraphics[width=0.442\textwidth]{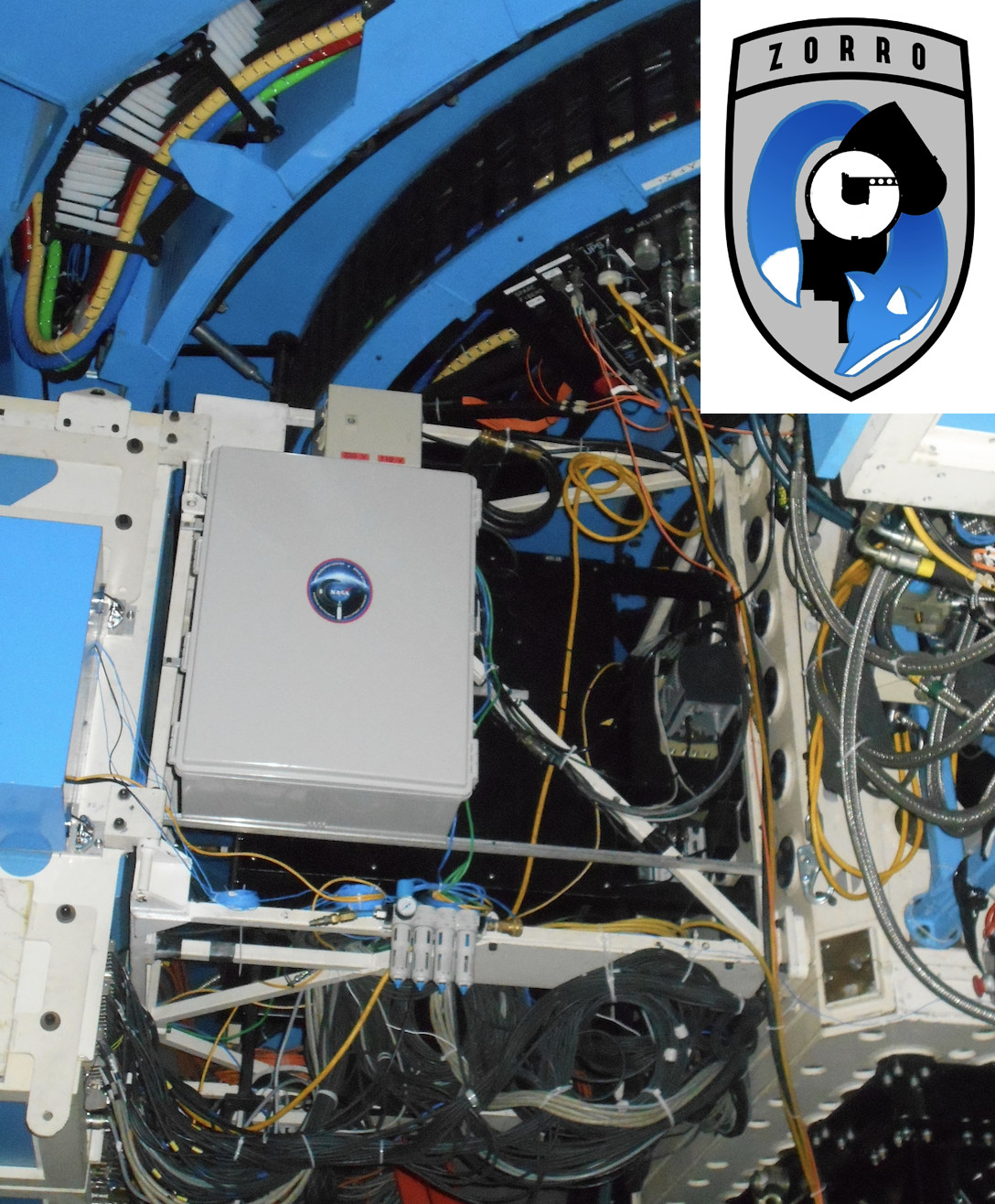}
\caption{\textit{Left:} The `Alopeke main instrument components were visible during a routine maintenance inspection at the Gemini North base facility. The ANDOR EMCCD cameras are seen extending from the main enclosure. \textit{Right:} Zorro mounted on the Gemini South telescope. The black box with one ANDOR camera seen extending from it contains the instrument, while the larger white box to the left contains the electronic power supplies and instrument computer.}\label{fig:alopeke&zorro}
\centering
\end{figure*}

`Alopeke and Zorro (the `$\rm{\overline o}$lelo Hawai'i and Spanish words for ``fox") are identical instruments that use iXon Ultra 888 EMCCD cameras to provide simultaneous speckle imaging in two optical band passes, yielding high-resolution reconstructed images of the observed source. SDSS u,g,r,i,z, H$\alpha$, and 4 narrow-band filters are available in two filter wheels passing light dichorically split at 700 nm. These imagers are used with one of two circular fields-of-view: speckle mode (diameter=$6.7$ arcseconds) or wide-field mode (diameter=$60$ arcseconds).  The left panel of Figure~\ref{fig:alopeke&zorro} shows `Alopeke with the covers removed, revealing the tightly packed innards that contain two filter wheels, the optical elements, and the two ANDOR EMCCD cameras extending from the box. The right panel of Figure~\ref{fig:alopeke&zorro} shows Zorro in its permanent mount location at the GCAL port, attached underneath the Gemini South 8.1-meter primary mirror. 
The small space available for the speckle instruments did not allow for an atmospheric dispersion corrector (ADC) to be included. However, this exclusion is offset by the large advantage that our visitor instruments have in that they are permanently mounted on Gemini and therefore always available to use. We discuss our ``no ADC" mitigation strategy below. A complete description of these two visiting Gemini instruments, including relative transmission as a function of wavelength in the 350-1000 nm range, can be found in \citet{scott:2021} and at the Gemini Observatory instrument web pages \footnote{https://www.gemini.edu/instrumentation/alopeke-zorro}.

`Alopeke and Zorro proposal demand at Gemini varies semester by semester, but on average it is 5-10\%, exceeding the Gemini South Adaptive Optics Imager (GSAOI) demand and nearing the demand for other visiting instruments (e.g., IGRINS), or even facility instruments (e.g., GNIRS). Any user can request time on these instruments using the regular Call for Proposals, or Director's Discretionary Time (DDT) when needed. Principal Investigators (PIs) from Gemini partner countries can also propose for `Alopeke and Zorro time using Fast Turnaround (FT) proposals \footnote{\url{https://www.gemini.edu/observing/schedules-and-queue/}}. 

\section{Observations, Data Reduction, and Image Reconstruction}

Speckle instrumentation is fairly simple, inexpensive, and small compared to other instruments mounted on large-aperture telescopes. `Alopeke and Zorro are roughly the size of a carry-on suitcase. The most difficult aspect of speckle observing is the quantity and short duration of the exposures, which is easily managed by the fast read-out capabilities of the EMCCDs. 
Using the Gemini mirror coating reflectivity, the optics and filter transmissions, and the QE of the EMCCDs, optimal speckle observations can be carried out between 400-940 nm where the total throughput is 40\% or higher \citep{scott:2021}.
As a mitigation for the lack of an ADC, typical speckle observations are performed at an airmass of 1.4-1.5 or lower (elevations above 45 degrees) and narrow band filters are used. Speckle imaging at Gemini is often done under (bright) moonlit sky conditions, for which the narrow band filters are preferred as well. The EMCCDs can also function as normal CCDs, allowing traditional, long exposure digital images to be obtained.

Speckle imaging of a target requires many thousands of short exposures (10 to 60 milliseconds in length) to be obtained and processed. This large number of images is required to build up sufficient S/N, especially at contrasts greater than four or five magnitudes ($\sim$10$^{-2}$), at very close angular separations, and/or for fainter targets.  Although this large number of exposures may seem daunting, at tens of milliseconds per image, typical speckle observations last only a few minutes per target \citep{hope:2019, 2022FrASS...9.1163H}. Stars with $V$ magnitudes from 1 to $\sim 12$ require only $\sim$5 minutes of observation time, during which 3,000 to 5,000 thousand exposures are collected. However, targets as faint as $R=19$ can be observed by using $\sim$ 50 minutes of on-source time. In a typical night, 40 to 50 sources are observed. \citet{scott:2021} and \citet{2022FrASS...9.1163H} provide full details on the relation between the magnitude of the source, the filter, the brightness of the target and the sky conditions to the total integration time at the source and the resulting S/N. To provide the reader with a gauge for the time required for a variety of observation types, we list in Section \ref{sec:Panoply}, the total on-source times and filters used for each observation. 

Almost all currently used speckle image reconstruction software packages are based on Fourier speckle interferometric methods \citep[e.g.,][]{labeyrie:1970,lohmann:1983,horch:2001,horch:2015}. Our standard data reduction pipeline \citep{horch:2012,howell:2011} provides robust $5\sigma$ magnitude contrast limits on stellar companion or circumstellar material detections \citep[e.g., ][]{howell:2016}.
The results presented in this paper, and all fully reduced data in the archives, are based on our implementation of the data reduction methods as described in \cite{howell:2011} and \cite{horch:2012}. Speckle reconstructions scale the output image maximum value to 1.0 at the center of the image. All of the images presented in this paper (except Nova V906 Car and Eros which are shown using just a red intensity color map) have their brightest pixel scaled to 1.0 and are presented on a min-max log scale using the MATLAB ``jet" colormap (Figure 2).

\begin{figure}
\hspace{-0.5cm}
\centering
\includegraphics[width=0.5\columnwidth]{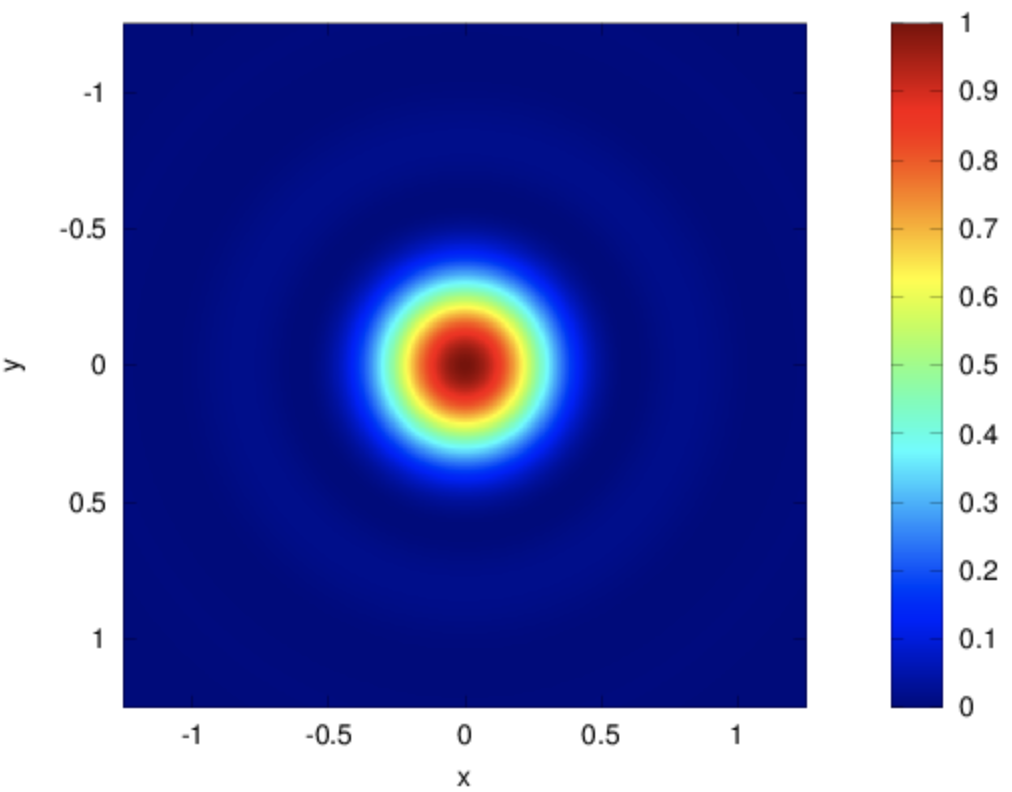}
\caption{The MATLAB ``jet" color map used for the images presented in this paper. }\label{fig:jet}
\centering
\end{figure}

Fourier speckle deconvolution techniques are computationally efficient but require a PSF model power spectrum usually obtained via routine observation of stars taken from the HR or HD catalogs that are known or assumed to be single based on past spectroscopic or imaging observations. These PSF standards are assigned to each target by the observing team prior to the scheduled observations. They are observed near in time to the targets and with similar sky locations. Each PSF standard observation requires an additional $\sim3$ minutes of observing time. Modern image reconstruction methods that are based on blind deconvolution techniques and can reach deeper contrasts plus provide more accurate astrophysical results are beginning to be implemented for speckle image reconstructions. \citep[e.g.,][]{2024AJ....167..258H}.

\begin{figure}
\centering
\hspace{-0.5cm}
\includegraphics[width=0.6\columnwidth]{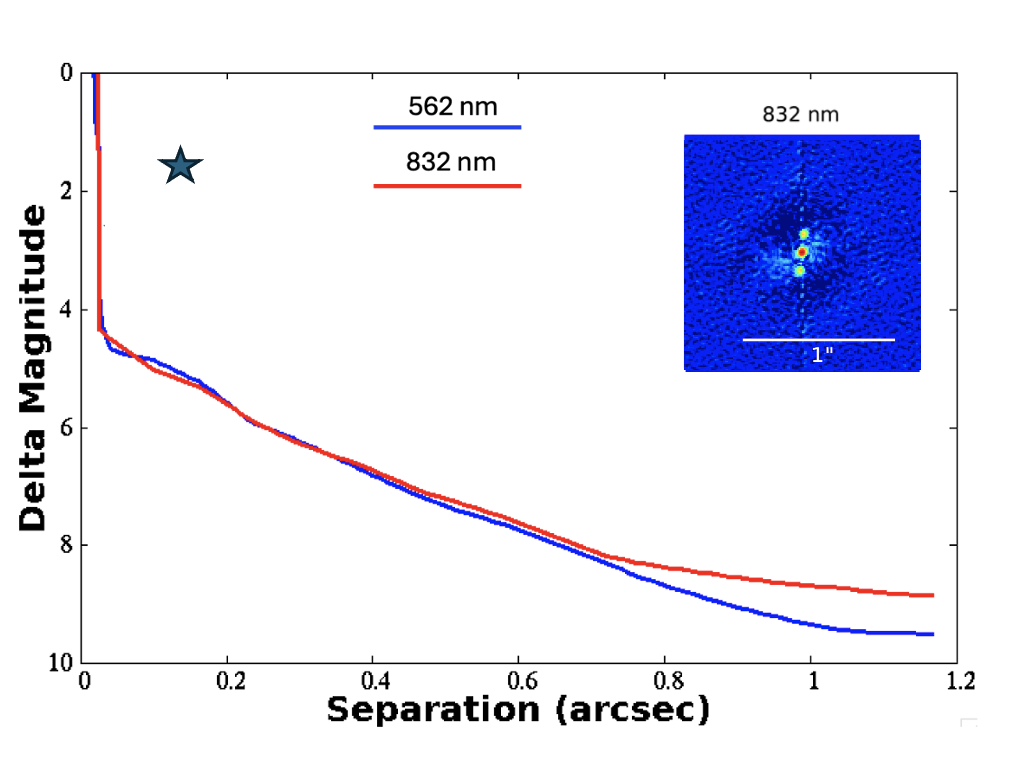}
\caption{A typical speckle imaging result for an 8th magnitude star (TOI-5873). The red and blue curves show the $5\sigma$ contrasts achieved as a function of angular separation, from the diffraction limit out to 1.2 arcseconds. The star symbol shows the properties of the companion, and the inset shows the reconstructed image for the 832~nm observation; a scale bar is included for reference. The speckle observations revealed that the target star is binary, with a very close (0.14 arcsecond separation) companion that is 1.6 magnitudes fainter than the primary at position angle 176 degrees. See \S3 for details.} \label{fig:brightstar}
\centering
\end{figure}

Figure~\ref{fig:brightstar} presents a typical Gemini speckle imaging result for a point source (the star TOI-5873), and the discovery of a very close stellar companion that is 1.6 magnitudes fainter than the primary. Note the 180-degree autocorrelation quadrant ambiguity that occurs when stars have comparable brightness \citep{horch:2012}. During data reduction, a ``best determination", using all sets of images taken of the target plus bispectrum analysis, is made to assign the correct location. 
The astrometric precision for binary stars observed at Gemini yields stellar separations to $\pm$1 mas and $\pm$1 degree in position angle \citep{lester:2021}.
Photometric magnitude differences for stars in multiple systems typically have uncertainties of $\pm$0.25 magnitudes, increasing by $\sim 2 \times$ for very close or wide ($>$0.9") pairs \citep[e.g.,][]{howell:2011,horch:2012}.
Raw and fully reduced archival data from Zorro and `Alopeke can be found in the Gemini Observatory \citep[][]{hirst:2017}\footnote{\url{https://archive.gemini.edu/searchform/ZORRO}, \\ \url{https://archive.gemini.edu/searchform/ALOPEKE}} or Exoplanet Follow-up Observing Program\footnote{\url{https://exofop.ipac.caltech.edu/tess/}} archives.

\section{A Panoply of Scientific Applications and Results} \label{sec:Panoply}

In this section, we discuss various scientific targets that have been observed using observations from `Alopeke and/or Zorro. Most of these observations were obtained as test cases or engineering studies (unless otherwise noted) and as such, many have not been published previously. These studies allowed researchers to assess the potential for using speckle imaging to accomplish their scientific goals, and have formed the basis for a number of stand-alone proposals targeting the detailed study of single objects, or larger sample collection to enable more global results.
 
\subsection{Stars}

Stars have been the predominant targets of the speckle imaging campaigns. In this section, we explore the varied results from these observations.

\subsubsection{Exoplanet-Hosting Stars}

\begin{figure}
\centering
\includegraphics[width=0.4\columnwidth]{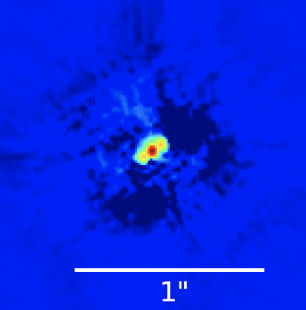}
\caption{The microlens source Kojima-1Lb observed with `Alopeke at Gemini North. This SDSS $i$ band image was obtained using 35 minutes of on-source time and detected both the lens and source separated by 0.058 arcsec with a magnitude difference of 3.7. The very close (blended in the reconstructed image) companion shows a 180-degree ambiguity with the true source location being at P.A.=110 degrees. North is up and East to the left in the image.}\label{fig:lens}
\centering
\end{figure}

Several large observing programs aimed at surveying exoplanet host stars have been in place at Gemini for many years. These programs have targeted transiting exoplanet candidates from Kepler, K2, TESS, and other NASA missions and have observed many thousands of stars \citep[e.g.,][]{lester:2021,howell:2021c,Review-2021FrASS...8...10H,Clark2022AJ....163..232C,matson2025AJ....169...76M}. Such surveys aim to identify close-in stellar companions that either induce false positive signals or contaminate the light curves of planets, leading to underestimated planetary radii \citep{Ciardi-2015ApJ...805...16C,Furlan-2017AJ....154...66F}. Figure~\ref{fig:brightstar} is an example of an 8th magnitude TESS exoplanet candidate host star observed at Gemini South with Zorro. Observations of exoplanet host stars continue for missions such as TESS, Roman, Ariel, and the Habitable Worlds Observatory. As the Roman space telescope prepares for launch, microlensing observations will reach a new high as numerous free-floating planets will be discovered via observations pointed toward the Galactic bulge. Speckle observations will play an important role in the resolution of the target star lens, as was done for the R$\sim$18 Kojima-1Lb microlens event (Figure~\ref{fig:lens}).

\subsubsection{Stellar Multiplicity}

\begin{figure*}
\centering
\includegraphics[width=0.278\textwidth, angle=0]{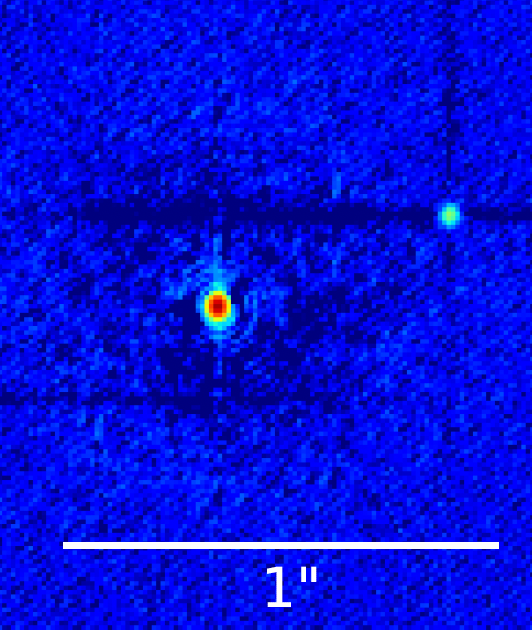}
\includegraphics[width=0.3\textwidth, angle=0]{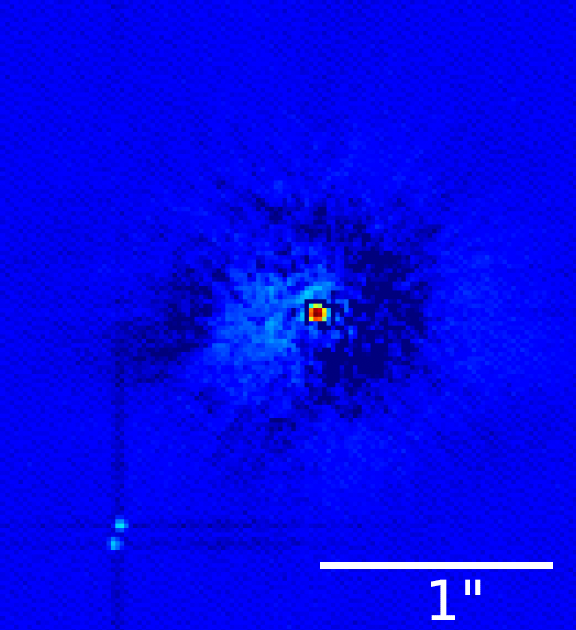}
\caption{\textit{Left:} The late M star WISE J133734.15-371100.2 as observed by Zorro. This 9th magnitude star was observed for 30 minutes at 832~nm. Note the close companion at a separation of 0.6 arcsec that is 3 magnitudes fainter. At a distance of 34~pc, the two stars are 20~au apart and have masses of 0.23 and 0.09~M$_{\odot}$ respectively. \textit{Right:} The planet hosting star TOI-697 was discovered to be triple using Gemini speckle observations. The primary star harbors a 2.6 R$_{\oplus}$ planet in a 8.6 day orbit and is itself orbited by a close binary pair 5 magnitudes fainter at a separation of 1.15 arcsec \citep{ARM2025MNRAS.537.3175A}. This 832~nm observation used 5 minutes of Gemini time.}\label{fig:M+L}
\centering
\end{figure*}

Surveys investigating the multiplicity of various types of stars have been carried out at the Gemini Observatory. These include low-mass stars \citep[e.g.,][]{Winters-2021AJ....161...63W,Clark2022AJ....163..232C}, halo stars, and higher-order multiplicity in known and metal-poor binaries \citep[e.g.,][]{zach-2022ApJ...934...72H,RENE2025arXiv250308721M}. Figure~\ref{fig:M+L} shows a faint likely L-dwarf companion to an M star identified by \citet{Deacon2007A&A...468..163D}, as well as a newly discovered, planet-hosting triple star system TOI-697 \citep{lester:2021}. High-resolution imaging of triple star systems has been proposed as a robust test of modified gravity \citep{REL2023OJAp....6E...2M}.

The space missions Kepler/K2 and TESS were designed to detect transiting exoplanets, but have also catalyzed a wealth of astrophysical discoveries. These findings include the detection of very high multiplicity stellar systems whose orbital planes are all edge-on to our line-of-sight. The speckle instruments at Gemini Observatory identified component stars in some of these doubly eclipsing quadruple systems and triply eclipsing triple systems \citep{Multi2024ApJ...974...25K}.

\subsubsection{Angular Diameters}

\begin{figure}
\centering
\vspace{0.3cm}
\includegraphics[width=0.4\columnwidth]{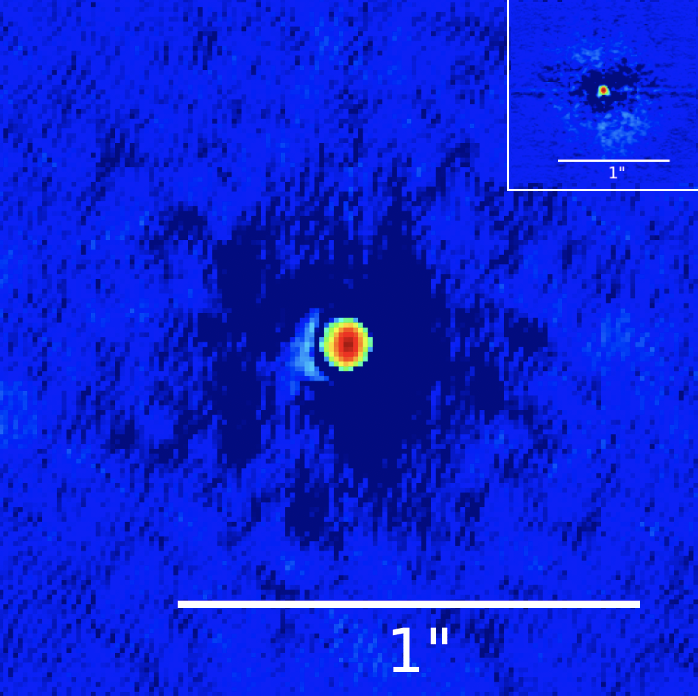}
\caption{A 562~nm `Alopeke observation of Betelgeuse obtained in February 2020, during the ``Great Dimming Event.'' Betelgeuse, $V=1.4$, is resolved with an angular diameter of 40~mas. The inset shows an unresolved diffraction limited PSF reference star observed near in time to Betelgeuse. This observation took a total of 12 minutes of Gemini time.}\label{fig:betel}
\centering
\end{figure}

Because speckle imaging allows the diffraction limit of the telescopes to be achieved, stars with large angular diameters can be resolved. `Alopeke on Gemini North was used to observe Betelgeuse during its ``Great Dimming Event" in February 2020. At the time, Betelgeuse had a $V$ magnitude of 1.4. Figure~\ref{fig:betel} shows a 562~nm speckle image of the star along with an unresolved PSF reference star observed near in time (Howell et al., 2025, in prep.). The disk of Betelgeuse is resolved with an apparent angular diameter of 40~mas.

\subsubsection{Stellar Eclipses}

\begin{figure}
\centering
\includegraphics[width=0.5\columnwidth]{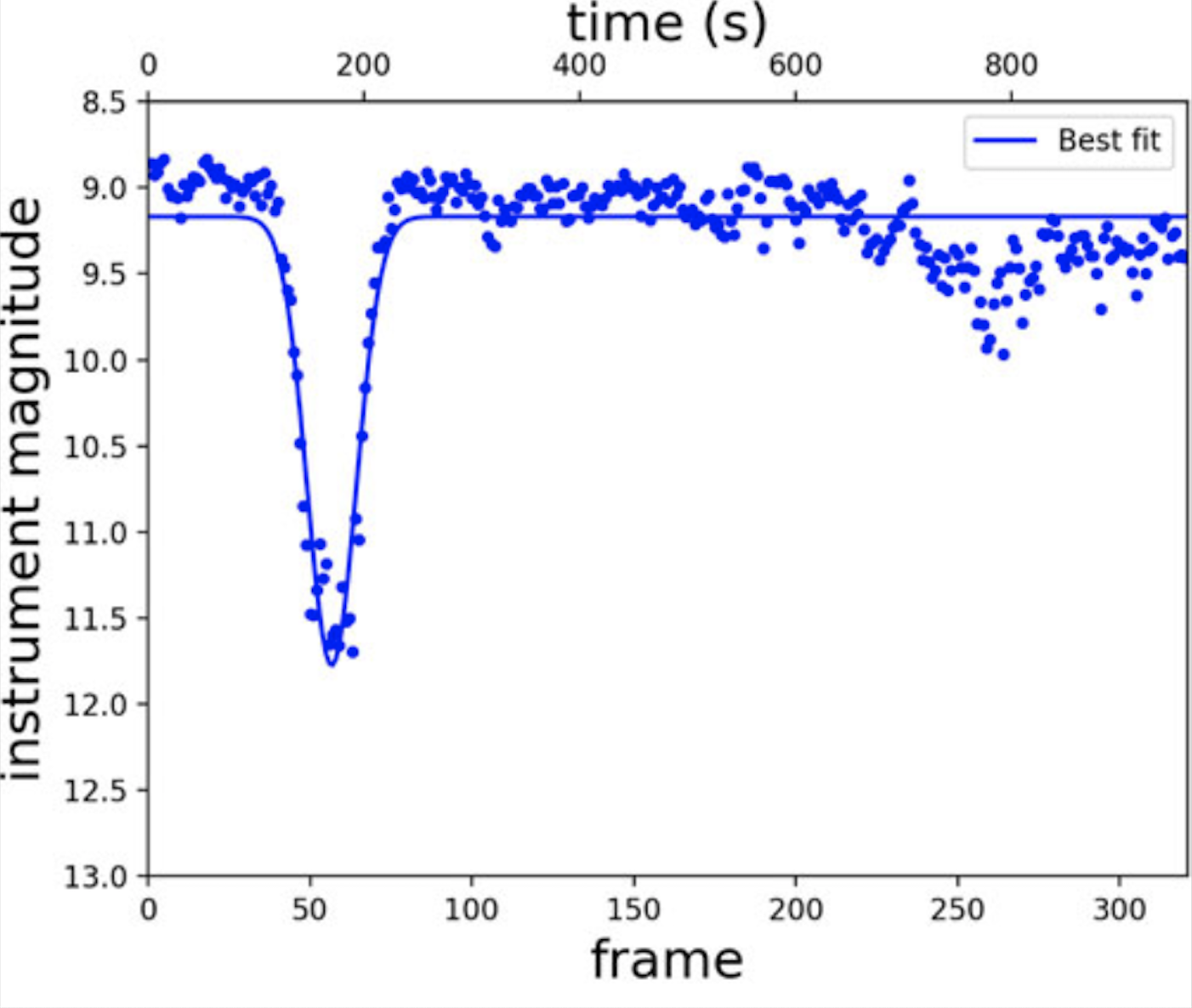}
\caption{Photometric eclipse light curve in SDSS i of the double degenerate binary star SDSS J0822+3048. The time series consisted of 2 hours of 15 second exposures that were used to provide ingress and egress details for the short eclipse. Adapted from \citet{scott:2021}.
\label{fig:ztf}}
\centering
\end{figure}

Figure~\ref{fig:ztf} shows a SDSS i light curve of the r=18 eclipsing double degenerate binary star SDSS J0822+3048 with a period of 40.5 minutes \citep{2021MNRAS.500.5098K}. Time series standard CCD observations, comprised of 15-second exposures, were simultaneously obtained in SDSS r and SDSS i for about 2 hours, allowing the very short, 90-second duration eclipse to be examined in detail.

\subsubsection{Evolved Stars and Stellar Remnants}

As stars end their lives, they often eject material into space either as soft puffs of atmospheric material or the rapid energetic explosion of a supernova. These sources often consist of a central point-source-like object surrounded by symmetric or very asymmetric material outflows \citep[see e.g.,][]{Huang2023ApJ...947...11H}. Evolved stars -- such as interacting binary systems, common envelope pairs, and symbiotic stars -- can often show resolved features as well.

\begin{figure*}
\centering
\includegraphics[width=0.49\textwidth]{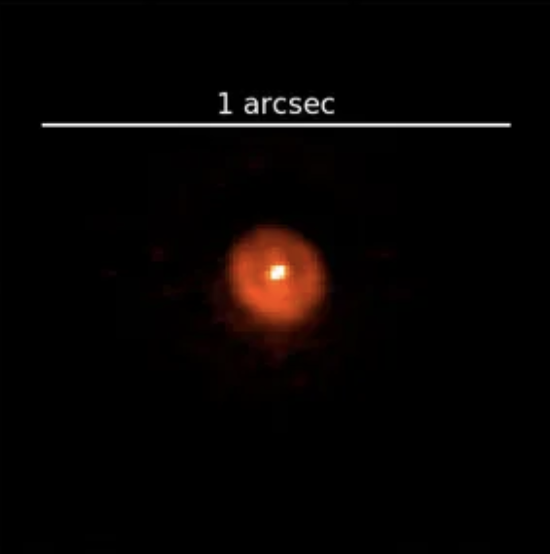}
\includegraphics[width=0.49\textwidth]{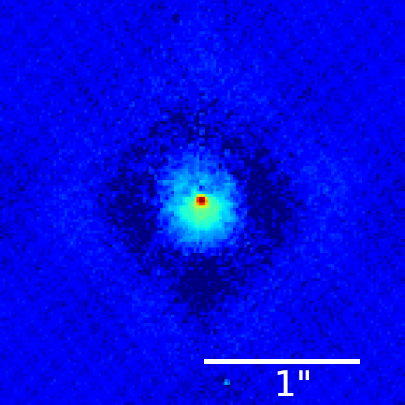}
\caption{Speckle images of resolved nova shells. \textit{Left:} Nova V906 Car observed in November 2020, 978 days after its explosion. The image was taken at 832~nm and shows a resolved nova shell with a radius of 90~mas. \textit{Right:} The classical nova V603 Aql (Nova Aql 1918) shows an asymmetric shell at 832~nm. Each of these observations used 20 minutes of Gemini time.} \label{fig:nova}
\centering
\end{figure*}

\begin{figure*}
\centering
\includegraphics[width=\textwidth]{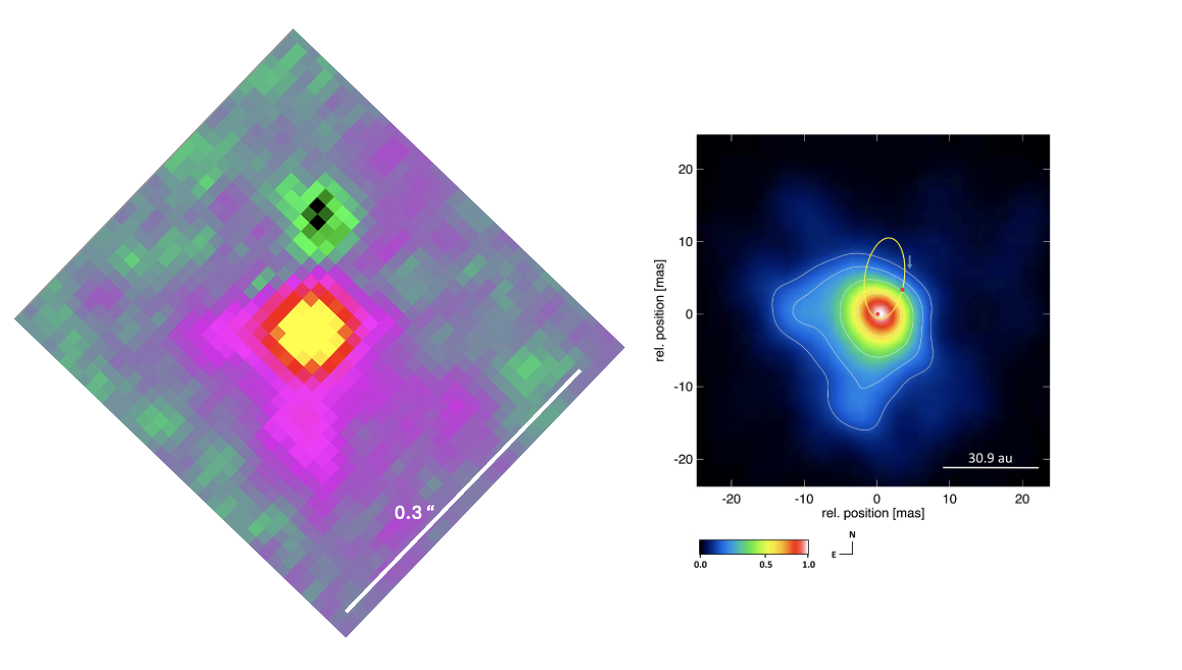}
\caption{Two images of the famous star Eta Carina. \textit{Left:} A false color composite composed of speckle images in the SDSS $u$, 562~nm, and 832~nm filters. These observations used a total of 40 minutes of Gemini time. The speckle image is $\sim$0.3 arcseconds (300 mas) on a side. \textit{Right:} The highest resolution image of Eta Car available today covering the inner 50 mas around the binary. This image was made using the ESO VLTI AMBER instrument in the K band \citep{Wei2016A&A...594A.106W}.
The orbit of the secondary star around the primary is shown for comparison.
The VLTI image contours outline the K band continuum wind emission which is seen to extend to larger distances in the optical speckle composite image.}\label{fig:etacar}
\centering
\end{figure*}

Speckle observations have allowed the measurement of winds from Wolf-Rayet stars \citep{shara-2023MNRAS.525.3195S}, shapes of stellar merger remnants \citep{zain-2024A&A...686A.260M}, nova shells (Figure~\ref{fig:nova}), supernovae \citep{van-2024ApJ...968...27V} and well-known specific targets such as R Aqr (Liimets et al. 2025, in prep.) and Eta Carinae (Figure~\ref{fig:etacar}). Eta Car was first observed with speckle imaging by \citet{Wei1986A&A...163L...5W} who found four starlike components in their field of view of 0.8". The Gemini speckle image shown in (Figure~\ref{fig:etacar}) reveals the central light concentration and fans of extended wind emission coming from the binary.

\subsection{Solar System Bodies}

Speckle imaging is also useful for objects in our own backyard. This section presents observations of Solar System objects.

\subsubsection{Angular Sizes}

\begin{figure*}
\centering
\includegraphics[width=0.2\textwidth]{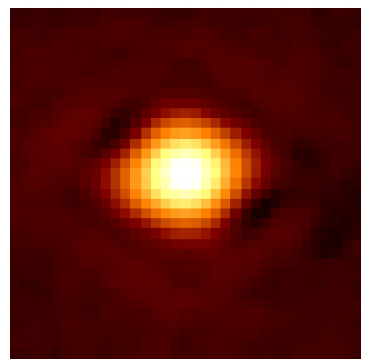}\includegraphics[width=0.2\textwidth]{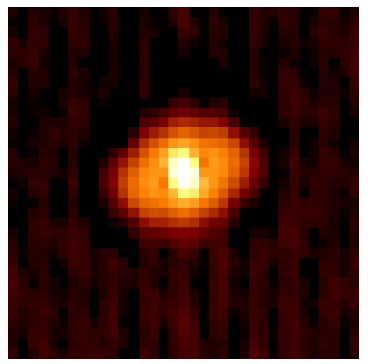}\includegraphics[width=0.2\textwidth]{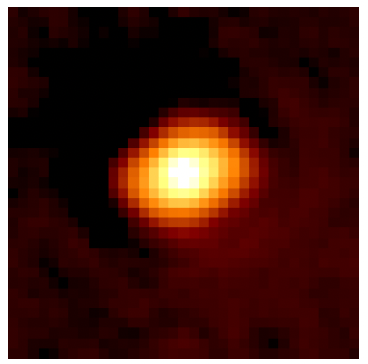}\includegraphics[width=0.2\textwidth]{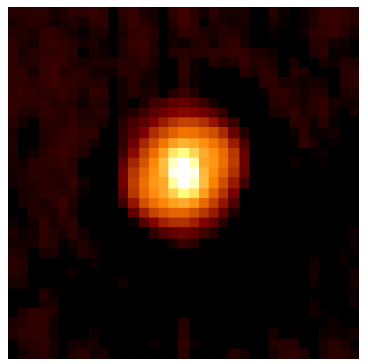}\includegraphics[width=0.2\textwidth]{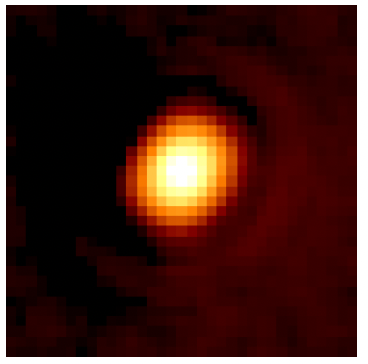}
\caption{Selected frames from a speckle imaging time series of the asteroid Eros that covered roughly one rotation period (5.3 hours). These 832~nm data were obtained in February 2020 at the Gemini South telescope using Zorro, and the time series data were reduced as 25 reconstructed images, each using five sets of 1000$\times$60~millisecond exposures and requiring $\sim$12 minutes of Gemini time. During the series, the apparent shape of the non-spherical body is seen changing with time.}\label{fig:eros}
\centering
\end{figure*}

Many objects in our Solar System are large enough to be resolved. Speckle images of Pluto and Charon provided measurements of their diameters, and were the highest resolution images of the (dwarf) planets until the New~Horizons fly-by \citep{Pluto2012PASP..124.1124H}. Asteroids have also been targets of the Gemini speckle imagers. During its close approach, the asteroid Phaethon (1983TB) was found to be 59~mas in angular size, corresponding to a physical size of 4.1~km \citep{2018LPI....49.1919W}. \citet{Q2023A&A...673L...4P} observed a stellar occultation by the Trans-Neptunian object Quaoar with the aim of improving its shape model and physical parameters and searching for additional material around the body. Figure~\ref{fig:eros} shows five frames from a 5.3-hour speckle imaging observation sequence of the asteroid 433 Eros. Eros is the second largest near-Earth asteroid. It has an elongated shape and a volume equivalent diameter of $\sim$17~km. The time series covered a single rotation period (P=5.3 hr) of the asteroid, and the apparent shape of the non-spherical body is seen to change with time. These data are unpublished and available in the Gemini archive.

\subsubsection{Occultations}

Pluto, asteroids, Kuiper Belt Objects, and other Solar System bodies have also had occultations observed by the Gemini speckle imagers, using simultaneous high-speed two-color photometry \citep[e.g.,][]{PL2023DPS....5530802S}. Other studies have observed target stars in advance of an occultation to assess whether they are single or multiple \citep[e.g.,][]{V2019AGUFM.P42C..08S}.

\subsection{Extragalactic Sources and Eruptive Events}

This last section presents several Gemini speckle imaging programs related to celestial objects that ``go bang in the night" and require fast action to observe the start of their eruptions, and/or fast sampling to reveal rapidly evolving structures in their light curves.

\subsubsection{Crowded Fields}

\begin{figure*}
\centering
\includegraphics[width=\textwidth]{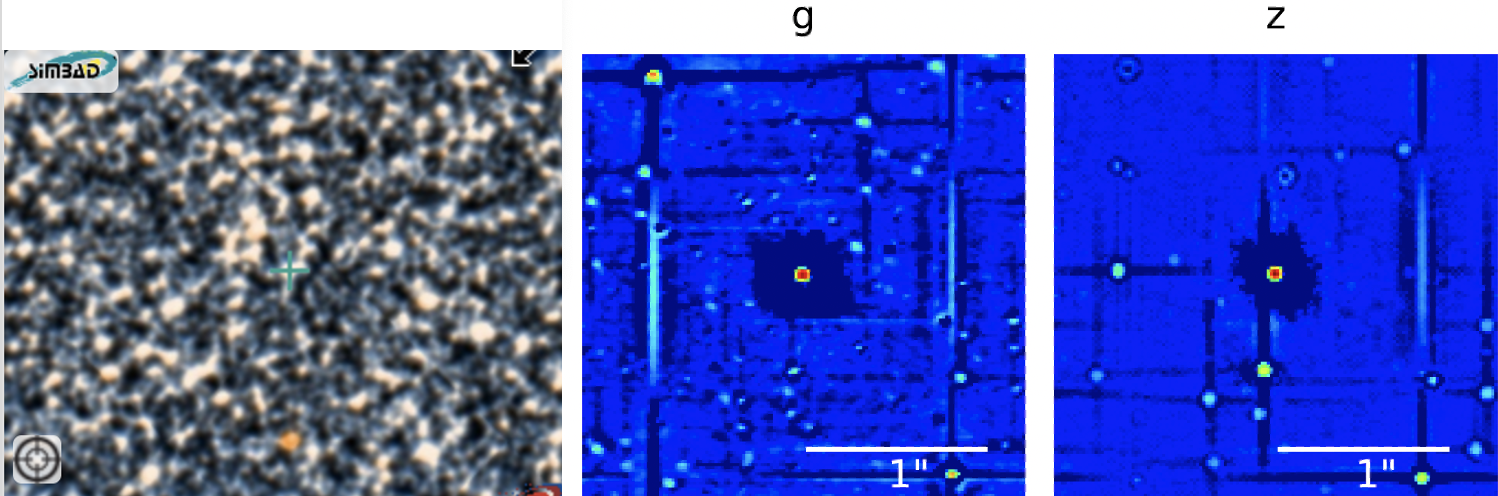}
\caption{V4641 Sgr, a $V\sim14$, high-mass black hole binary, is easily isolated from its very crowded field in speckle images. \textit{Left:} A DSS image showing a $4\times4$ arcminute region around V4641 Sgr (from SIMBAD). \textit{Right:} Simultaneous SDSS $g$ and SDSS $z$ speckle images for which photometry can be performed without issues of crowding. These images were obtained using `Alopeke and 12 minutes of Gemini time.}\label{fig:chart}
\centering
\end{figure*}

The black hole binary V4641 Sgr is in an extremely crowded field (Figure~\ref{fig:chart}). Speckle observations provided not only a precise location of the source, but unblended photometric measurements as well. Speckle imaging has also been used in crowded regions of the Magellanic Clouds to search for binaries and resolve the core of R136 \citep{Kal12022ApJ...935..162K,Kal22024ApJ...972....3K}.

\subsubsection{Dual Quasars}

Observation of a 19th magnitude dual quasar candidate resulted in the detection of both binary black hole nuclei, and required only 50 minutes of Gemini time \citep{QSO2021RNAAS...5..210H}. Earlier speckle interferometry imaging observations for quasars, even with 6-m telescopes, were limited to brighter ($\le$16) targets. New speckle observations with Gemini allow nearly diffraction-limited optical imaging opening a new venue to the confirmation and detailed studies of high-z dual quasars.

\subsubsection{X-ray Binary Outbursts}

Outbursts of X-ray binaries offer opportunities to measure quasi-periodic oscillations during the accretion process, yielding insights into the underlying disks. \citet{XRB2022cosp...44.1750T} used speckle imaging to perform an optical fast timing study of various X-ray binaries. The simultaneous two-color data collected enabled the examination of disk energetics and jet physics. Furthermore, \citet{scott:2018} showed, using engineering data from the WIYN telescope,  that the ability to detect both pulse shape and pulse period in two optical colors at the same time allows precise identification of the pulsational modes involved in the pulsating white dwarf HL Tau.

\subsubsection{Fast Radio Bursts}

Fast radio bursts (FRB) are mysterious transient sources. FRB 20180916B repeats with a known period, making it a useful test case. \citet{FRB2024ApJ...964..121K} used `Alopeke to observe FRB 20180916B simultaneously in the SDSS $r$ and SDSS $i$ bands, requiring 20 minutes of Gemini time. No optical burst was detected at the time of the radio burst, allowing certain FRB models to be ruled out.

\section{Summary}

The `Alopeke and Zorro speckle cameras are permanent visitor instruments at the Gemini North and South telescopes in Hawai'i and Chile. One can request to use these instruments for general queue proposals, DDT and FT proposals, Long and Large Proposals, and Target of Opportunity proposals as well. Note that the oversubscription rate of the Gemini telescopes hovers near 3, as compared to 6-7 for HST, Chandra, and JWST.

Optical speckle imaging has the advantage of darker skies (than the IR), no need for AO natural or laser guide stars, inexpensive and simple instrumentation, and easy setup and use. Gemini speckle observations can be made from 350~nm to 1000~nm, providing the ability to gather a spectral energy distribution across the optical bandpass for any detected companion. Additionally, speckle imaging reaches the diffraction limit and is indifferent to whether the target star is single or multiple; the observations and data reduction processes are uniformly applied regardless of the target's multiplicity. This is not an ability shared by coronagraphic instruments.

The applications of speckle imaging have broadened greatly since `Alopeke and Zorro were first installed at Gemini Observatory. While first used to observe stars at high angular resolution to search for stellar companions, speckle imaging is now employed to decipher the shape of Solar System bodies, study the outflows and ejected material of supernova, observe extragalactic eruptive events, and investigate other exciting astrophysical phenomena.

As the world of ground-based observational astronomy moves into the era of 30-40 meter-class telescopes, speckle imagers should be one of the first facility instruments available. Speckle imagers are inexpensive, simple to construct, small in format, light in weight, and easy to operate.
The use of EMCCDs and their high-speed imaging ability not only allows speckle imaging to be accomplished but fast time-series photometric observations as well.
Such time-series observations of bright stars can also find telescope engineering applications in the analysis of x,y centroids over time to examine guider and tracking errors, and unwanted telescope vibration modes. This type of engineering study has already been successfully implimented at Gemini using 'Alopeke and Zorro. 
The high-speed imaging cameras can also be used as telescope guiders or wavefront sensors, such as has been done at the WIYN telescope for use with the NEID spectrograph \citep[see e.g.,][]{Gupta2021AJ....161..130G}.

Speckle instrument imagers can also function as standard CCD imagers. Providing, as we have in 'Alopeke and Zorro, an optional wider field of view, the EMCCD cameras can be used as conventional CCD imagers, providing long-exposure photometric images. As a ``first-light" instrument, obtaining images throughout the larger final field of view would yield optical quality checks of the delivered PSF across the focal plane. 

As speckle interferometry moves to these larger telescopes, reaching the optical diffraction limit on a 30-meter-class telescope would provide, at 400~nm, an spectacular ground-based angular resolution of 4~mas. Such resolution would be unprecedented, allowing a multitude of new astrophysical science cases to be explored.

\section*{Author Contributions}
All authors listed have made a substantial, direct and intellectual contribution to the work, and approved it for publication.
SH \& CM-V: Writing – original draft, review and editing. EF, NS, RM, CL, CC, KL, ZH, DC, \& SD: review and editing.

\section*{Conflict Of Interest}
The authors declare that the research was conducted in the absence of any commercial or financial relationships that could be construed as a potential conflict of interest.

\section*{Generative AI statement}
The authors declare that no Generative AI was used in the creation of this manuscript.

\section*{Acknowledgements}
We would like to acknowledge the staff at the international Gemini Observatory for their many hours of support, help, and friendship that allowed this high-resolution imaging to be possible. Science Operation Specialists, contact scientists, and queue coordinators have been particularly supportive and patient with us during our programs. We also appreciate the allocations of engineering time that allowed us to test various observational modes and increase community service.
An especially big thanks to Mark Everett, Dave Mills, Rebecca Gore, Sergio Fajardo-Acosta, Andy Adamson, Jeong-Eun Heo, Atsuko Nitta, Fredrik Rantakyro,  Joanna Thomas-Osip, Venu Kalari, Ricardo Salinas, Andrew Stephens, and John White. 

The authors also thank the American Astronomical Society for organizing meetings with poster and free-time community gatherings that promote friendly and productive conversations that can turn into interesting research and publications such as this paper.

The observations in this paper made use of the High-Resolution Imaging instruments `Alopeke and Zorro. `Alopeke and Zorro were funded by the NASA Exoplanet Exploration Program and built at the NASA Ames Research Center by Steve B. Howell, Nic Scott, Elliott P.~Horch, and Emmett Quigley. `Alopeke and Zorro are mounted on both 8.1-m telescopes of the international Gemini Observatory, a program of NSF NOIRLab, which is managed by the Association of Universities for Research in Astronomy (AURA) under a cooperative agreement with the U.S. National Science Foundation, on behalf of the Gemini partnership: the National Science Foundation (United States), National Research Council (Canada), Agencia Nacional de Investigación y Desarrollo (Chile), Ministerio de Ciencia, Tecnología e Innovación (Argentina), Ministério da Ciência, Tecnologia, Inovações e Comunicações (Brazil), and Korea Astronomy and Space Science Institute (Republic of Korea).

This research has made use of the NASA Exoplanet Archive and ExoFOP, which are operated by the California Institute of Technology, under contract with the National Aeronautics and Space Administration under the Exoplanet Exploration Program. Additional information was obtained from the SIMBAD database, operated at CDS, Strasbourg, France. We acknowledge support from AFOSR awards FA9550-14-1-0178 (DAH and SMJ) and FA9550-21-1-0384 (SMJ).

\bigskip
\noindent {\it {Facilities:}} Gemini -  `Alopeke, Zorro

\clearpage

\bibliographystyle{Frontiers-Harvard.bst}
%\bibliography{Speckle.bib}
 \newcommand{\noop}[1]{}

\end{document}